\RequirePackage{fix-cm}
\documentclass[twocolumn,epjc3]{svjour3}  
\smartqed  
\RequirePackage{graphicx}
\usepackage{amsmath}
\usepackage{cite}

\usepackage{multicol}

\usepackage{lipsum}
\usepackage{cuted}

\journalname{Journal}
\begin{document}

\title{A two-field dark energy model with cubic contractions of the Riemann tensor
}

\author{Mihai Marciu\thanksref{e1,addr1}
}

\thankstext{e1}{e-mail: mihai.marciu@drd.unibuc.ro}

\institute{Faculty of Physics, University of Bucharest, Bucharest-Magurele, Romania\label{addr1}
}

\date{Received: date / Accepted: date}

\maketitle

\begin{abstract}
The paper proposes a novel cosmological model based on a two--field scenario, where the two fields are independently coupled with a specific invariant, based on cubic contractions of the Riemann tensor. After obtaining the modified Friedmann equations and the specific Klein--Gordon relations, the investigation studies the corresponding physical features by adopting the linear stability theory in a quintom scenario case where the coupling functions and potential energies have an exponential behavior. The corresponding investigation revealed a high complexity of the phase space structure, with various stationary points which can explain different aspects in the evolution of our Universe, the matter dominated epoch, and the late time accelerated stage. The present model can represent a viable solution to the dark energy problem due to the high complexity of the phase space and the existence of the scaling solutions. 
\end{abstract}

\section{Introduction}
\label{intro}
\par 
In the present days the technological advances permit the study of the history and evolution of the Universe in a more profound manner, leading to a series of questions and breakthroughs from a scientific point of view. The exploration of various questions related to the history of the Universe can open new extraordinary possibilities in the long run. On a short time scale, the quest for the major cosmological scientific questions pose new specific difficulties to various technological areas, leading to the development of new technologies in applied science. The discovery of the accelerated expansion of the current Universe was a breakthrough at the end of the last millennium \cite{Peebles:2002gy}, configuring new specific areas in physics. This new curious phenomenon has been probed through various phenomenological studies \cite{DES:2021wwk, Planck:2015fie}, representing a viable and intriguing aspect of modern physics \cite{Beutler:2011hx}. The simplest dark energy model is represented by the $\Lambda$CDM scenario where the accelerated expansion is driven by a cosmological constant \cite{Copeland:2006wr}. However, this model has various limitations and cannot explain various observed features, including the late time dynamical evolution of the dark energy equation of state \cite{Copeland:2006wr}. In order to overcome these issues, a new theoretical direction has emerged, the extended theories of gravity \cite{Capozziello:2011et}. These theories are based on the mathematical approach of basic general relativity, extending the fundamental action to a more general or special form \cite{Clifton:2011jh, Bamba:2012cp}. It can be seen that the modified gravity theories can explain various aspects from the evolution of the Universe \cite{Nojiri:2017ncd}. In these theories, the fundamental action is extended, being replaced with a generic functional which takes into account possible influences from various invariant geometrical quantities \cite{Koyama:2015vza}. Another viable approach consists in adding one or more scalar fields in the fundamental action \cite{Tsujikawa:2010zza, Kasper:1989wu}, relating the evolution of the late time acceleration of the Universe with the dynamics induced by the scalar field(s). The scalar fields which can trigger the acceleration of the Universe can be in the form of quintessence models, with a canonical kinetic energy \cite{Nojiri:2006ri, Tsujikawa:2013fta}, embedding also a potential term. Another approach where the scalar fields are non-canonical is represented by phantom dark energy models, an approach which can explain the super--acceleration \cite{Caldwell:2003vq, Vikman:2004dc, Nojiri:2005sx, Ludwick:2017tox}. Soon after the emergence of quintessence and phantom models, a new intriguing aspect was discovered -- the crossing of the cosmological constant boundary in the recent past for the dark energy equation of state \cite{Feng:2004ad, Huterer:2004ch, Nesseris:2005ur}. This phenomenon triggered the formation of quintom models \cite{MohseniSadjadi:2006hb, Alimohammadi:2006tw, Guo:2006pc, Shi:2008df, Wei:2007rp, Lazkoz:2007mx, Leon:2018lnd}, a complex construction which includes two scalar fields, one with a negative kinetic energy, and another, with a canonical kinetic energy. Such a construction can explain the crossing of the phantom divide line in the recent past, being extended also in the non--minimal case \cite{Marciu:2020yaw,Marciu:2020vve,Marciu:2020hpk,Marciu:2019cpb, Marciu:2018oks, Bahamonde:2018miw, Marciu:2016aqq}.   
\par 
In the scalar tensor theories of gravitation the Einsteinian cubic gravity was developed in Ref.~\cite{Bueno:2016xff}, representing a particular extension which includes higher order contractions of the Riemann tensor. Such an approach has attracted attention in the last couple of years, being studied in various cosmological applications \cite{Bueno:2018yzo, Bueno:2018xqc, Jiang:2019kks, Pookkillath:2020iqq, Bueno:2016ypa, Li:2017txk, Hennigar:2018hza, Bueno:2019ltp, Caceres:2020jrf, BeltranJimenez:2020lee, Edelstein:2022xlb, Rudra:2022qbv}. The extension towards a generic theory based on the cubic invariant has been proposed in \cite{Erices:2019mkd}, analyzed numerically. Later, such an approach has been studied \cite{Marciu:2020ysf, Quiros:2020uhr} dynamically considering the linear stability theory, investigating the physical features of the phase space structure. The inclusion of the cubic component for scalar fields has appeared shortly \cite{Marciu:2020ski, Marciu:2022wzh} for tachyonic models and quintessence or phantom components. Furthermore, a more general extension which takes into account also the scalar curvature has been investigated \cite{Marciu:2021rdl}. The observational analysis for the generic cubic gravity \cite{Giri:2021amc} showed the regions of interest for various parameters from an astrophysical point of view. Moreover, various authors have investigated the inflation \cite{Quiros:2020eim, Edelstein:2020nhg, Arciniega:2019oxa, Arciniega:2018tnn, Arciniega:2018fxj} and the black hole solutions \cite{Hennigar:2016gkm, Bueno:2016lrh, Feng:2017tev, Adair:2020vso} in different cubic gravitational theories. 
\par 
As previously stated, the two--field dark energy models in the form of quintom scenarios can explain the dynamical evolution of the dark energy equation of state and the possible crossing of the phantom divide line in the near past \cite{Cai:2009zp}. This aspect is exhibited in the minimal coupling case, where the fields have only kinetic and potential energies, as well as in the non--minimal models which take into account various geometrical invariant components in the specific action \cite{Cai:2009zp}. To this regard, the basic two--field dark energy models can in principle be extended, by considering particular invariant components based on third order contractions of the Riemann tensor. 
\par 
In the present paper we have developed a two--field dark energy model which takes into account possible couplings with a specific invariant obtained by applying special contractions of the Riemann tensor in the third order. Hence, the Einstein--Hilbert action is extended by adding two scalar fields independently coupled with an invariant based on cubic contractions of the Riemann tensor. After proposing the fundamental action for the cosmological model we obtain the field equations by varying the action with respect to the inverse metric in the case of a Roberson--Walker background. The resulting Klein--Gordon relations are obtained in the usual manner, by the variation with respect to the specific fields which are assumed to be time--depending. The physical features for this cosmological setup are investigated by applying the linear stability theory in the case of exponential coupling functions, determining the phase space structure and the corresponding stationary points which are associated to different eras in the evolution of the Universe. The dynamical characteristics for our model are investigated by analyzing the corresponding eigenvalues, obtaining possible constraints for the specific parameters which are related to different dynamical effects. The present paper can be regarded as a more general dark energy model which takes into consideration viable couplings between scalar fields and cubic contractions of the Riemann tensor, aiming towards a multi--scalar field theory.
\par
The paper is organized as follows: in Sec.~\ref{sec:1} we propose the action for the present cosmological model and discuss the corresponding field equations which are obtained by taking the variation with respect to the inverse metric. Then, in Sec.~\ref{sec:2} we analyze the features of the phase space structure in the case of an exponential coupling and potential energy, considering a quintom scenario. Lastly, in Sec.~\ref{sec:3} we summarize our investigation and discuss the main physical features obtained. 

\section{The description of the action and the corresponding field equations}
\label{sec:1}

\par 
In this section we shall investigate a cosmological model based on two scalar fields which are non--minimally coupled with an invariant based on cubic contractions of the Riemann tensor, having the following action: 
\begin{multline}
\label{actiune}
S=S_m+\int d^4x \sqrt{-\tilde{g}} \Bigg( \frac{R}{2}-\frac{\epsilon_1}{2} \tilde{g}^{\mu\nu}\partial_{\mu}\phi\partial_{\nu}\phi-\frac{\epsilon_2}{2} \tilde{g}^{\mu\nu}\partial_{\mu}\sigma\partial_{\nu}\sigma
\\-V_1(\phi)- V_2(\sigma) + f(\phi)P + g(\sigma) P\Bigg),
\end{multline}
with the cubic invariant \cite{Erices:2019mkd}:
\begin{multline}
P=\beta_1 R_{\mu\quad\nu}^{\quad\rho\quad\sigma}R_{ \rho\quad\sigma}^{\quad \gamma\quad\delta}R_{\gamma\quad\delta}^{\quad\mu\quad\nu}+\beta_2 R_{\mu\nu}^{\rho\sigma}R_{\rho\sigma}^{\gamma\delta}R_{\gamma\delta}^{\mu\nu}
\\+\beta_3 R^{\sigma\gamma}R_{\mu\nu\rho\sigma}R_{\quad\quad\gamma}^{\mu\nu\rho}+\beta_4 R R_{\mu\nu\rho\sigma}R^{\mu\nu\rho\sigma}+\beta_5 R_{\mu\nu\rho\sigma}R^{\mu\rho}R^{\nu\sigma}
\\+\beta_6 R_{\mu}^{\nu}R_{\nu}^{\rho}R_{\rho}^{\mu}+\beta_7 R_{\mu\nu}R^{\mu\nu}R+\beta_8 R^3.
\end{multline}
\par 
As can be noted, we have extended the fundamental Einstein--Hilbert action by considering two additional scalar fields endowed with potential energy, non--minimally coupled with a specific term which encodes effects from the third order contractions of the Riemann tensor. In this action the two constants $\epsilon_{1,2}$ describe the signs of the kinetic terms and can describe the case where the two scalar fields are canonical or non--canonical. Before proceeding to the dynamics, we need to describe the background, by specifying the Robertson--Walker metric:
\begin{equation}
\label{metrica}
ds^2=-dt^2+a^2(t) \delta_{ju}dx^j dx^u,
\end{equation}
with $a(t)$ the cosmic scale factor, characterizing the accelerated expansion of the Universe, defining the Hubble parameter in the usual manner, $H=\dot{a}/a$. 
\par 
Next, by considering the following relations \cite{Erices:2019mkd}:
\begin{equation}
\beta_7=\frac{1}{12}\big[3\beta_1-24\beta_2-16\beta_3-48\beta_4-5\beta_5-9\beta_6\big],
\end{equation}
\begin{equation}
\beta_8=\frac{1}{72}\big[-6\beta_1+36\beta_2+22\beta_3+64\beta_4+5\beta_5+9\beta_6\big],
\end{equation}
\begin{equation}
\beta_6=4\beta_2+2\beta_3+8\beta_4+\beta_5,
\end{equation}
\begin{equation}
\beta=(-\beta_1+4\beta_2+2\beta_3+8\beta_4),
\end{equation}
we can show that the cubic invariant term for the Roberson--Walker metric takes the following form \cite{Erices:2019mkd}:
\begin{equation}
\label{PP}
P=6 \beta H^4 (2H^2+3\dot{H}).
\end{equation}
\par
The modified Friedmann equations which are associated to the cosmological model are obtained by considering the variation with respect to the inverse metric, reducing to the following \cite{Marciu:2020ski}:
\begin{equation}
\label{friedmannconstr}
3H^2=\rho_m+\rho_{\phi}+\rho_{\sigma},
\end{equation}
\begin{equation}
\label{friedmannaccelerare}
3H^2+2\dot{H}=-p_m-p_{\phi}-p_{\sigma},
\end{equation}
where 
\begin{equation}
\label{densitatede}
\rho_{\phi}=\frac{1}{2}\epsilon_1\dot{\phi}^2+V_1(\phi)+6 \beta f(\phi) H^6-18 \beta H^5 \frac{df(\phi)}{d\phi}\dot{\phi},
\end{equation}

\begin{equation}
\label{densitatede2}
\rho_{\sigma}=\frac{1}{2}\epsilon_2\dot{\sigma}^2+V_2(\sigma)+6 \beta g(\sigma) H^6-18 \beta H^5 \frac{dg(\sigma)}{d\sigma}\dot{\sigma},
\end{equation}

\begin{multline}
\label{presiunede}
p_{\phi}=\frac{1}{2}\epsilon\dot{\phi}^2-V_1(\phi)-6 \beta f(\phi) H^6-12 \beta f(\phi) H^4 \dot{H}
\\+12 \beta H^5 \frac{df(\phi)}{d\phi}\dot{\phi}+24 \beta H^3 \frac{df(\phi)}{d\phi}\dot{H}\dot{\phi}
\\
+6 \beta H^4 \dot{\phi}^2\frac{d^2f(\phi)}{d\phi^2}+6 \beta H^4 \frac{df(\phi)}{d\phi}\ddot{\phi},
\end{multline}

\begin{multline}
\label{presiunede2}
p_{\sigma}=\frac{1}{2}\epsilon_2\dot{\sigma}^2-V_2(\sigma)-6 \beta g(\sigma) H^6-12 \beta g(\sigma) H^4 \dot{H}
\\+12 \beta H^5 \frac{dg(\sigma)}{d\sigma}\dot{\sigma}+24 \beta H^3 \frac{dg(\sigma)}{d\sigma}\dot{H}\dot{\sigma}
\\
+6 \beta H^4 \dot{\sigma}^2\frac{d^2g(\sigma)}{d\sigma^2}+6 \beta H^4 \frac{dg(\sigma)}{d\sigma}\ddot{\sigma}.
\end{multline}

\par 
We can define the equation of state for the scalar fields:

\begin{equation}
w_{\bf{\phi\sigma}}=\frac{p_{\phi}+p_{\sigma}}{\rho_{\phi}+\rho_{\sigma}}.
\end{equation}

Then, the total equation of state is equal to:
\begin{equation}
w_{\bf{tot}}=\frac{p_m+p_{\phi}+p_{\sigma}}{\rho_{m}+\rho_{\phi}+\rho_{\sigma}}=-1-\frac{2}{3}\frac{\dot{H}}{H^2}.
\end{equation}
\par 
The Klein--Gordon equation for the first scalar field is obtained by taking the variation with respect to the $\phi$ field and reduces to \cite{Marciu:2020ski}:
\begin{equation}
\label{eq:eqkg1}
\epsilon_1(\ddot{\phi}+3 H \dot{\phi})+\frac{dV_1(\phi)}{d\phi}-6 \beta H^4 (2 H^2+3 \dot{H})\frac{df(\phi)}{d\phi}=0.
\end{equation}
\par 
In a similar way the second Klein--Gordon relation is obtained and has the following form:
\begin{equation}
\label{eq:eqkg2}
\epsilon_2(\ddot{\sigma}+3 H \dot{\sigma})+\frac{dV_2(\sigma)}{d\sigma}-6 \beta H^4 (2 H^2+3 \dot{H})\frac{dg(\sigma)}{d\sigma}=0.
\end{equation}

\par 
Finally, we can define the matter density parameter

\begin{equation}
\Omega_m=\frac{\rho_{m}}{3H^2},
\end{equation}
 and the density parameter associated to the present two field dark energy model:
\begin{equation}
\Omega_{de}=\frac{\rho_{\phi}+\rho_{\sigma}}{3H^2},
\end{equation}
satisfying the normal constraint:
\begin{equation}
\Omega_m+\Omega_{de}=1.
\end{equation}

\begin{figure}
  \includegraphics[width=8cm]{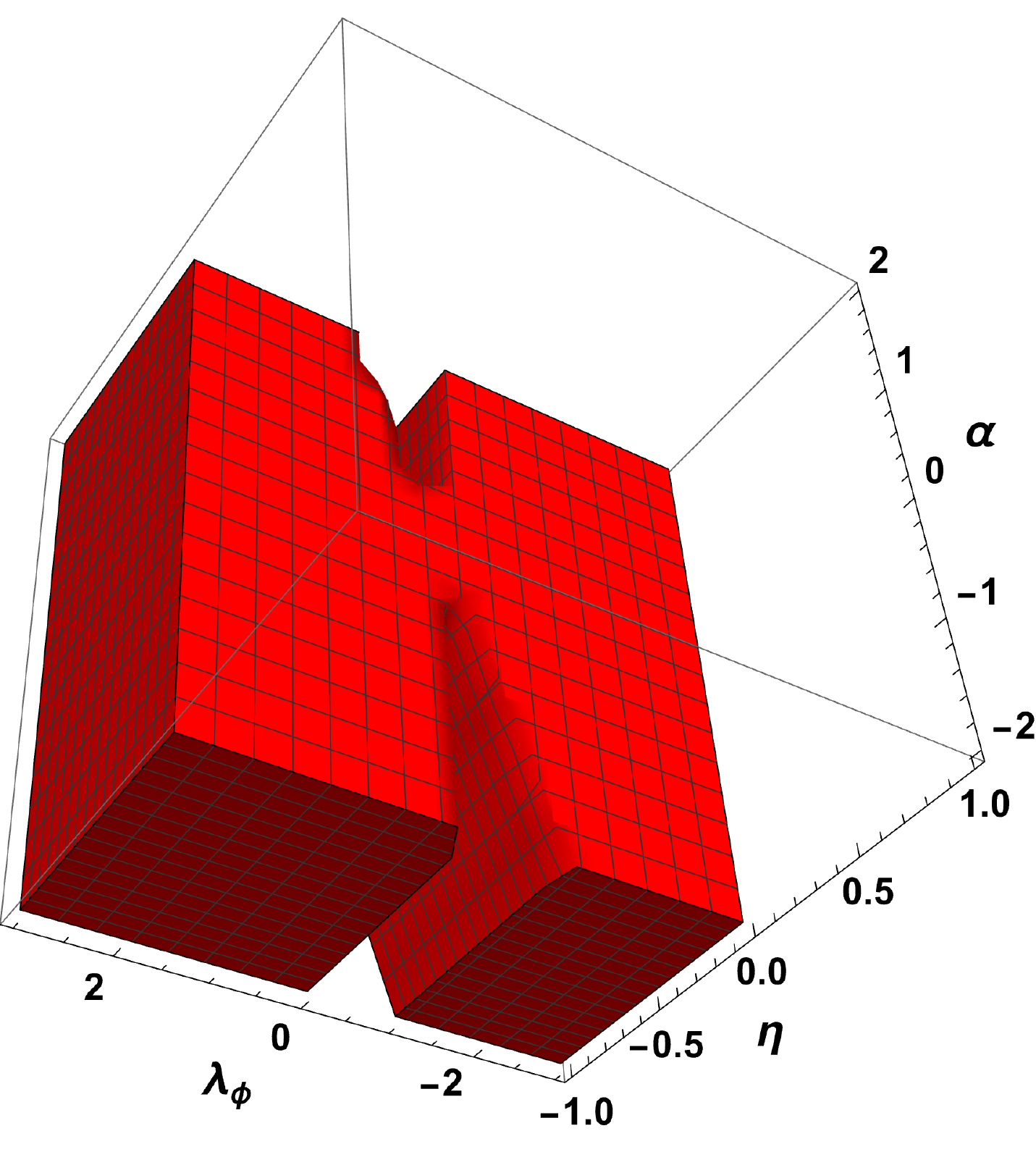}
\caption{The figure displays a possible saddle region for the $P_6$ critical point ($\lambda_{\sigma}=2$).}
\label{fig:1}       
\end{figure}

\begin{figure}
  \includegraphics[width=8cm]{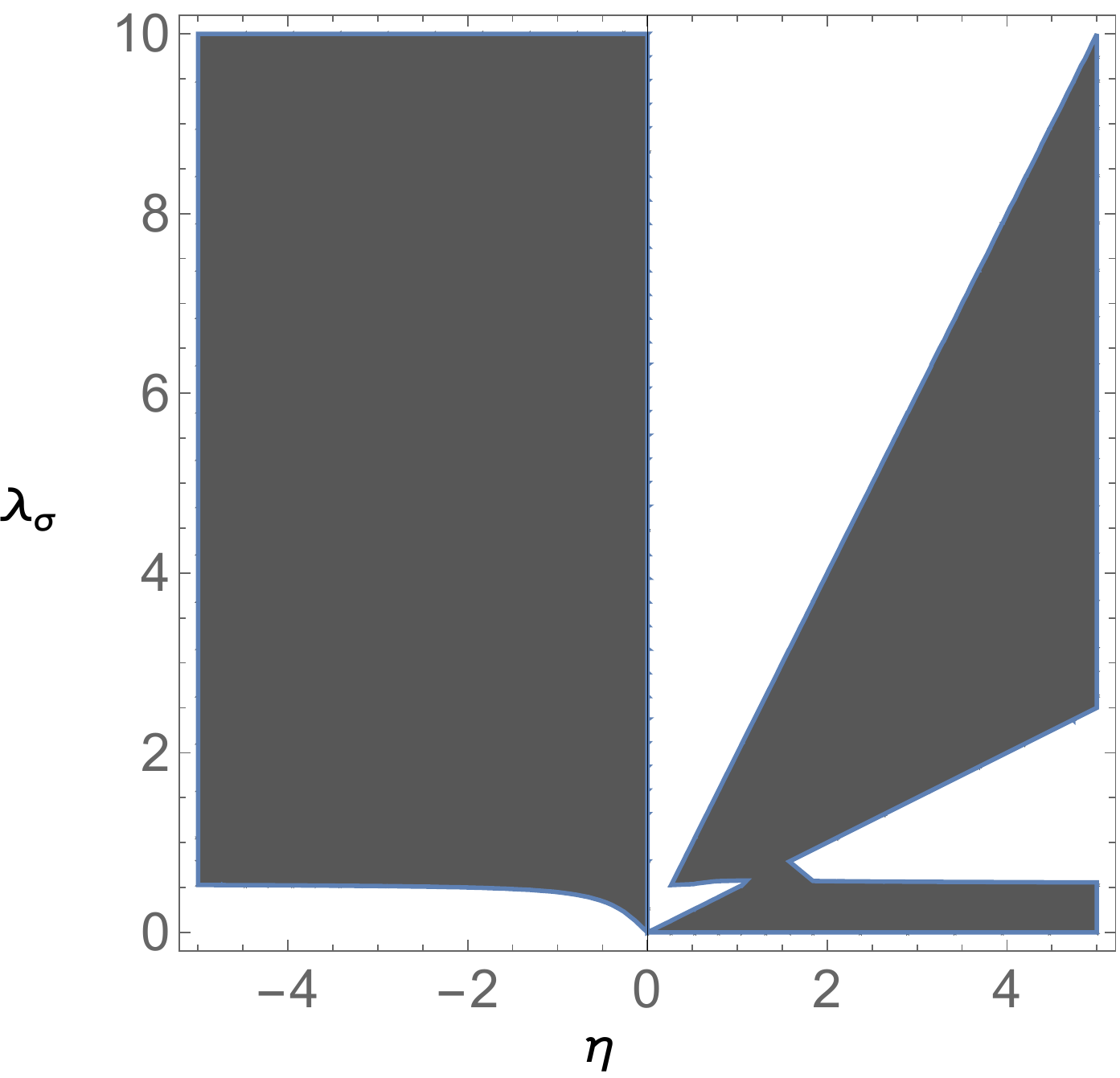}
\caption{A non--exclusive saddle region for the $P_7$ solution where the de--Sitter behavior is expected  ($w_m=0$).}
\label{fig:2}       
\end{figure}

\begin{figure}
  \includegraphics[width=8cm]{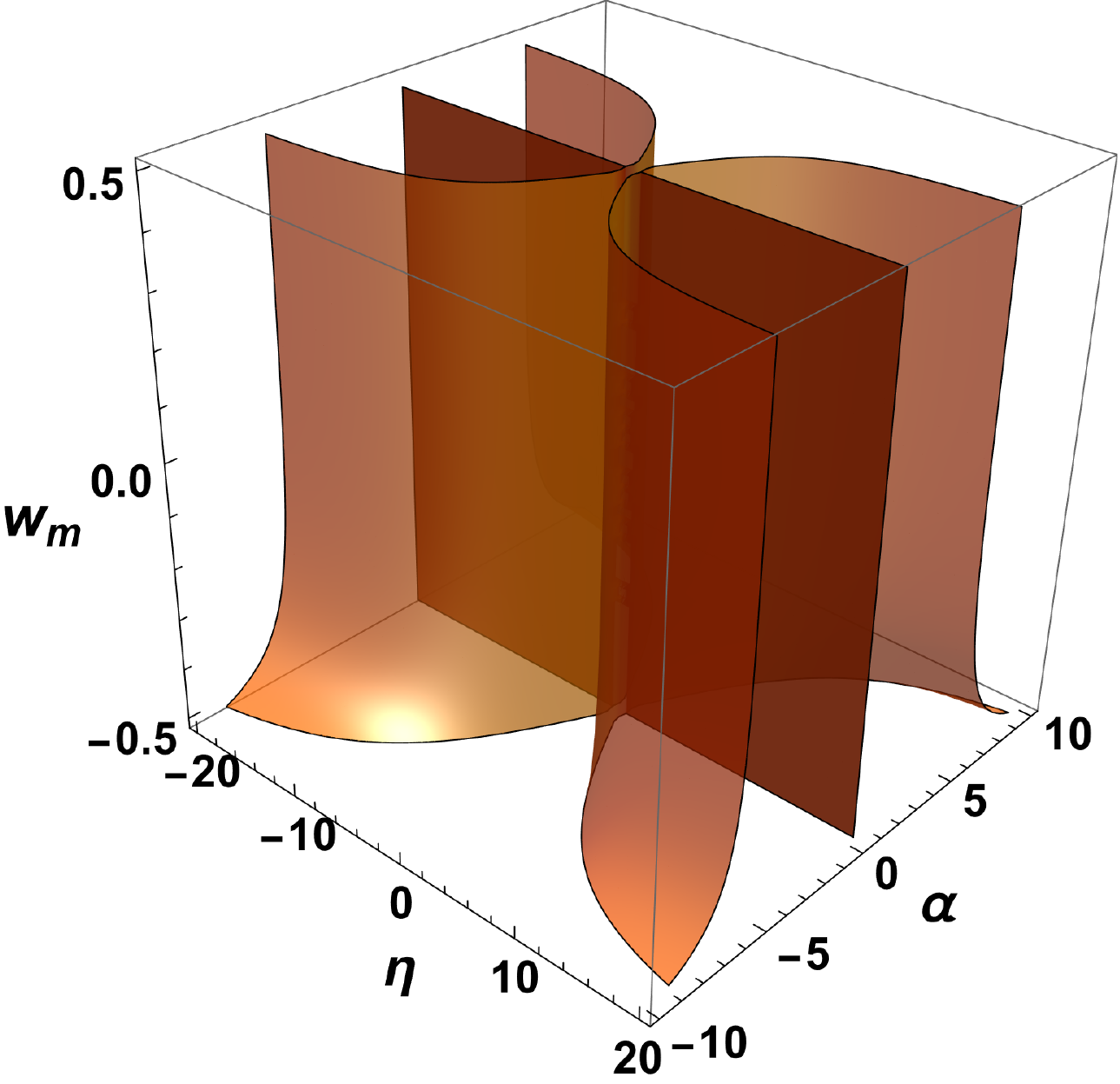}
\caption{A non--exclusive region for the $P_9$ solution where the matter density parameter is $s=0.3$.}
\label{fig:3}       
\end{figure}

\begin{figure}
  \includegraphics[width=8cm]{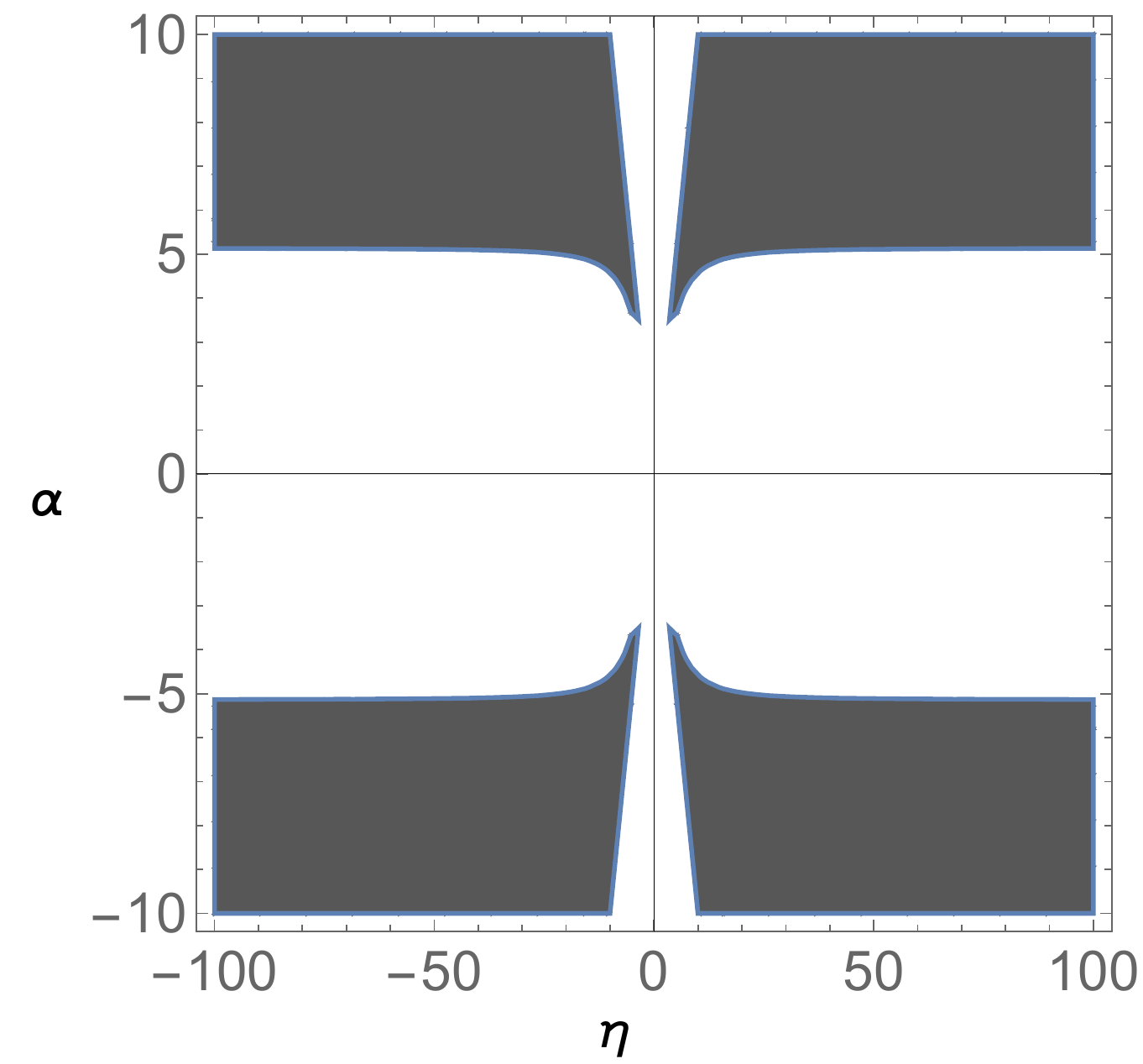}
\caption{A non--exclusive region where the $P_9$ solution is saddle ($w_m=0, \lambda_{\phi}=1, \lambda_{\sigma}=1$).}
\label{fig:4}       
\end{figure}

\begin{figure}
  \includegraphics[width=8cm]{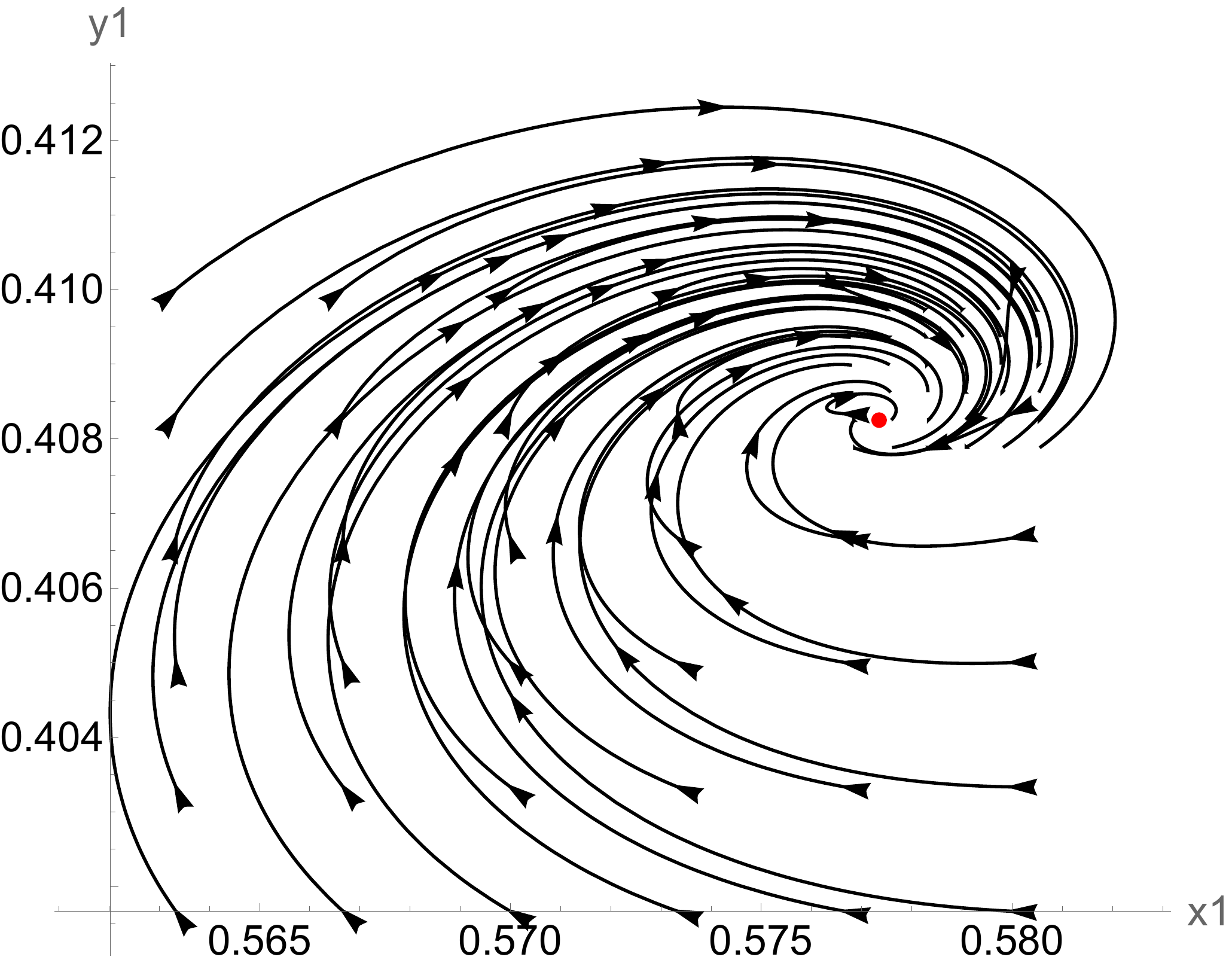}
\caption{The evolution near the $P_{11}$ critical point (represented as a red dot) in the $O x_1 y_1$ plane ($w_m=0, \lambda_{\phi}=3, \lambda_{\sigma}=1, \eta=1, \alpha=1$).}
\label{fig:p11}       
\end{figure}

\begin{figure}
  \includegraphics[width=8cm]{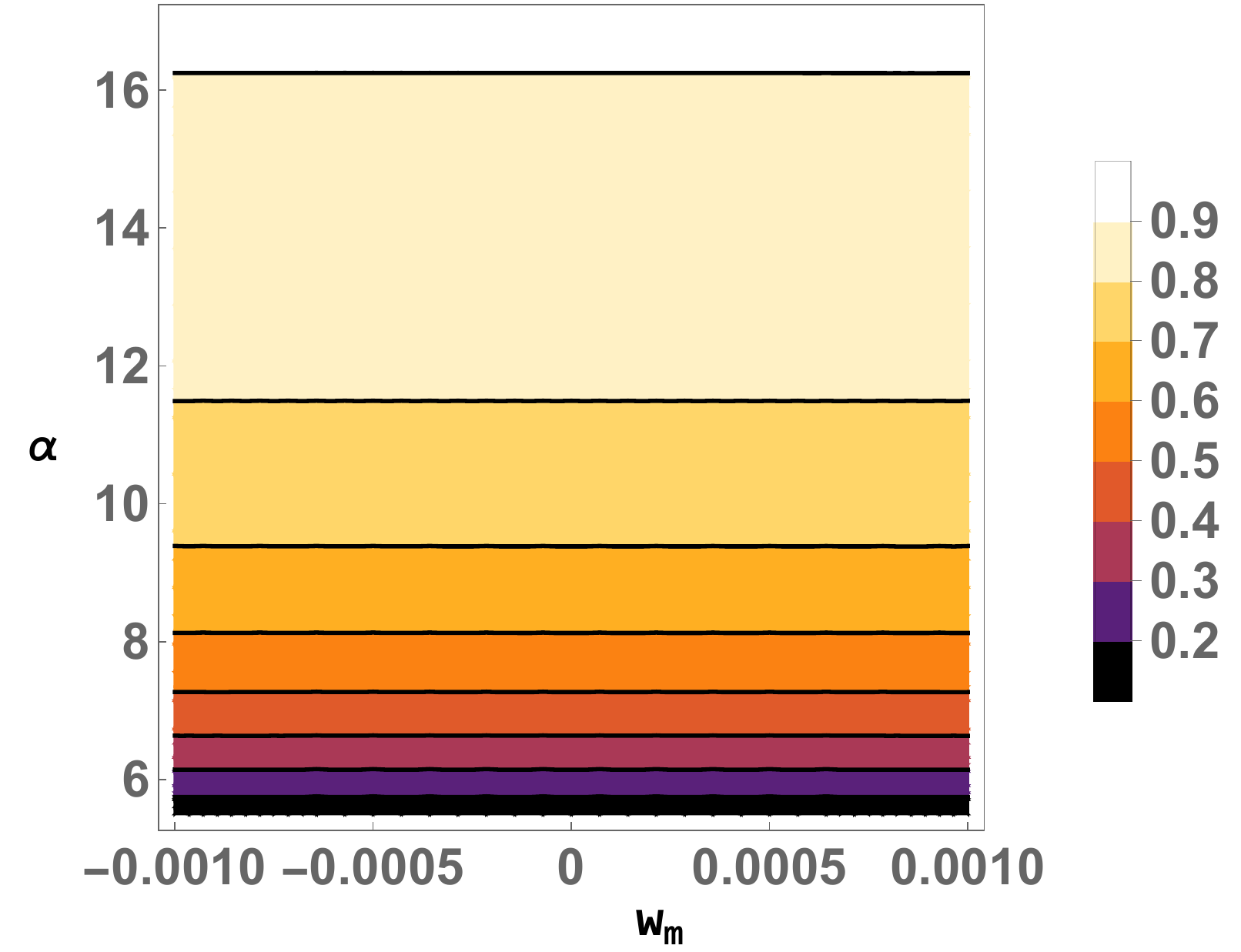}
\caption{The matter density parameter for the $P_{13}$ cosmological solution.}
\label{fig:5}       
\end{figure}

\begin{figure}
  \includegraphics[width=8cm]{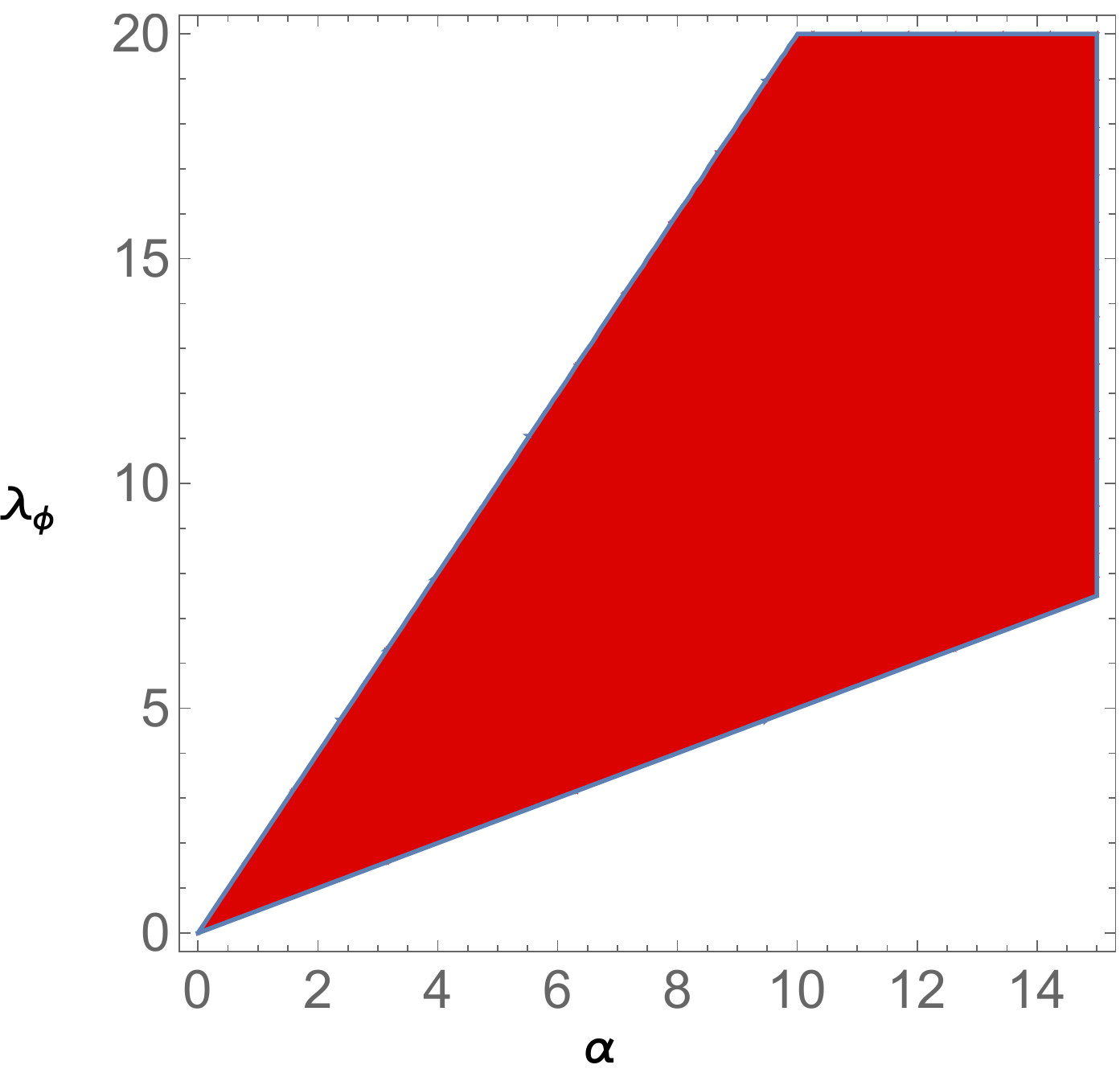}
\caption{A region of interest for the $P_{14}$ cosmological solution where the dynamics corresponds to a saddle behavior ($w_m=0$).}
\label{fig:6}       
\end{figure}

\section{Dynamical properties for a quintom scenario}
\label{sec:2}

In this section we shall investigate the dynamical properties of the present cosmological model by considering that we have a quintom scenario with $\epsilon_1=+1, \epsilon_2=-1$. In this case the $\phi$ component is associated to the quintessence, while the $\sigma$ term describes a non--canonical field with a negative kinetic energy violating various aspects from a physical point of view. The dynamical aspects are investigated by adopting the linear stability theory, an important tool in modern cosmology. In order to study the dynamical features, we introduce the following dimensionless variables: 

\begin{equation}
   x_1=\frac{\dot{\phi}}{H\sqrt{3}} ,
\end{equation}

\begin{equation}
   x_2=\frac{\dot{\sigma}}{H\sqrt{3}}, 
\end{equation}

\begin{equation}
   y_1=\frac{\sqrt{V_1(\phi)}}{H\sqrt{3}} ,
\end{equation}

\begin{equation}
   y_2=\frac{\sqrt{V_2(\sigma)}}{H\sqrt{3}} ,
\end{equation}

\begin{equation}
   z_1=2 \beta f(\phi) H^4 ,
\end{equation}

\begin{equation}
   z_2=2 \beta g(\sigma) H^4 ,
\end{equation}

\begin{equation}
   s=\frac{\rho_m}{3 H^2}.
\end{equation}

For the potential energy terms we shall consider the exponential decomposition, 

\begin{equation}
    V_1(\phi)=V_{10}e^{- \lambda_{\phi} \phi},
\end{equation}

\begin{equation}
    V_2(\sigma)=V_{20}e^{- \lambda_{\sigma} \sigma},
\end{equation}

where $V_{10}, V_{20}, \lambda_{\phi}$ and $\lambda_{\sigma}$ are constant positive parameters. Furthermore, for the coupling functions we have taken:

\begin{equation}
    f(\phi)=f_0 e^{\alpha \phi},
\end{equation}

\begin{equation}
    g(\sigma)=g_0 e^{\eta \sigma},
\end{equation}

with $f_0, g_0, \alpha, \eta$ constant parameters associated to the strength of the cubic dependence.

In order to study the physical features of the phase space structure, we shall introduce a new variable $N=log(a)$, approximating the evolution of our cosmological model by an autonomous system of differential equations. In this case we shall obtain the following autonomous system:

\begin{equation}
\label{eq:dn1}
    \frac{dx_1}{dN}=-x_1 \frac{\dot{H}}{H^2}+\frac{\ddot{\phi}}{\sqrt{3}H^2},
\end{equation}

\begin{equation}
    \frac{dx_2}{dN}=-x_2 \frac{\dot{H}}{H^2}+\frac{\ddot{\sigma}}{\sqrt{3}H^2},
\end{equation}

\begin{equation}
    \frac{dy_1}{dN}=-y_1 \frac{\dot{H}}{H^2}-\frac{\sqrt{3}}{2} \lambda_{\phi} x_1 y_1,
\end{equation}

\begin{equation}
    \frac{dy_2}{dN}=-y_2 \frac{\dot{H}}{H^2}-\frac{\sqrt{3}}{2} \lambda_{\sigma} x_2 y_2,
\end{equation}

\begin{equation}
    \frac{dz_1}{dN}=4 z_1 \frac{\dot{H}}{H^2}+\sqrt{3} \alpha x_1 z_1,
\end{equation}

\begin{equation}
\label{eq:dn2}
    \frac{dz_1}{dN}=4 z_2 \frac{\dot{H}}{H^2}+\sqrt{3} \eta x_2 z_2.
\end{equation}

Then, the Friedmann constraint equation reduces to:
\begin{equation}
    s=3 \sqrt{3} \alpha  x_1 z_1+3 \sqrt{3} \eta  x_2 z_2-\frac{x_1^2}{2}+\frac{x_2^2}{2}-y_1^2-y_2^2-z_1-z_2+1.
\end{equation}
In order to close the autonomous system we need to rewrite the Klein--Gordon equations in terms of auxiliary variables, obtaining:
\begin{equation}
    \ddot{\phi}=-3 \sqrt{3} H^2 x_1+3 H^2 y_1^2 \lambda _{\phi }+6 \alpha  H^2 z_1+9 \alpha  z_1 \dot{H},
\end{equation}

\begin{equation}
    \ddot{\sigma}=-3 \sqrt{3} H^2 x_2-3 H^2 y_2^2 \lambda _{\sigma }-6 \eta  H^2 z_2-9 \eta  z_2 \dot{H}.
\end{equation}

Lastly, the second Friedmann equation, the acceleration relation reduces to the following expression:

\begin{multline}
   -3 H^2-2 \dot{H}=3 H^2 s w_m+9 \alpha ^2 H^2 x_1^2 z_1+6 \sqrt{3} \alpha  H^2 x_1 z_1
   \\+9 \eta ^2 H^2 x_2^2 z_2+6 \sqrt{3} \eta  H^2 x_2 z_2+\frac{3}{2} H^2 x_1^2-\frac{3}{2} H^2 x_2^2-3 H^2 y_1^2
   \\-3 H^2 y_2^2-3 H^2 z_1-3 H^2 z_2+12 \sqrt{3} \alpha  x_1 z_1 \dot{H}+12 \sqrt{3} \eta  x_2 z_2 \dot{H}
   \\-6 z_1 \dot{H}-6 z_2 \dot{H}+3 \alpha  z_1 \ddot{\phi}+3 \eta  z_2 \ddot{\sigma},
\end{multline}
closing the system of differential equations in an autonomous manner. For the system of equations \eqref{eq:dn1}--\eqref{eq:dn2} we have identified various critical points which are associated to different cosmological epochs. In what follows we shall discuss the obtained results and the corresponding dynamical features. 
\par 
The first critical point represents the origin of the phase space structure, located at the following coordinates:
\begin{equation}
    O:=[x_1=0,x_2=0,y_1=0,y_2=0,z_1=0,z_2=0],
\end{equation}
where the total equation of state corresponds to a matter era ($w_{tot}=w_m, s=1$), with the following eigenvalues:
\begin{multline}
    \Big[ 
    \frac{3}{2} \left(w_m-1\right),\frac{3}{2} \left(w_m-1\right),-6 \left(w_m+1\right),-6 \left(w_m+1\right),
    \\ 
    \frac{3}{2} \left(w_m+1\right),\frac{3}{2} \left(w_m+1\right)
    \Big].
\end{multline}
In the case where $w_m \to 0$ the solution is associated to a saddle dynamical behavior, describing the late matter epoch.
\par 
The second solution is located to the following coordinates:
\begin{equation}
    P_1^{\pm}:=[x_2= \pm \sqrt{x_1^2-2},y_1= 0,y_2= 0,z_1= 0,z_2= 0],
\end{equation}
describing a critical line which is related to the kinetic variables of the two scalar fields. This solution appears also in the minimally coupled case, describing a stiff fluid solution ($w_{tot}=1, s=0$), a particular representation which is not of great interest nowadays. The corresponding eigenvalues are the following:
\begin{multline}
    \Big[ 
    0,3-3 w_m,\sqrt{3} \alpha  x_1-12,3-\frac{1}{2} \sqrt{3} x_1 \lambda _{\phi },\pm \sqrt{3} \eta  \sqrt{x_1^2-2}-12,
    \\ 
    \mp \frac{1}{2} \sqrt{3} \sqrt{x_1^2-2} \lambda _{\sigma }+3
    \Big].
\end{multline}
Since the second eigenvalue is always positive when  $w_m \to 0$ this solution cannot be stable, it is either saddle or unstable. Moreover, the solution is non-hyperbolic due to the existence of a zero eigenvalue. Due to this aspect, the linear stability theory can be used only to constrain and study the saddle characteristics, a feature which is particular for two-field dark energy models.
\par 
The third cosmological solution is represented by a de--Sitter ($w_{tot}=-1, s=0$) epoch located at the following coordinates:
\begin{multline}
    P_2:=[x_1=0,x_2=0,y_1=\frac{\sqrt{2} \sqrt{-\alpha } \sqrt{z_1}}{\sqrt{\lambda _{\phi }}},
    \\
    y_2=\frac{\sqrt{2} \sqrt{\eta  \lambda _{\phi }+2 \alpha  \eta  z_1-\eta  z_1 \lambda _{\phi }}}{\sqrt{2 \eta  \lambda _{\phi }-\lambda _{\sigma } \lambda _{\phi }}},
    \\
    z_2=\frac{-\lambda _{\sigma } \lambda _{\phi }-2 \alpha  z_1 \lambda _{\sigma }+z_1 \lambda _{\sigma } \lambda _{\phi }}{\lambda _{\phi } \left(2 \eta -\lambda _{\sigma }\right)}],
\end{multline}
describing a cosmological solution where the two dark energy fields are at rest without any kinetic energy. The dynamics is driven by the potential energy terms, and the strength of the cubic couplings embedded into $\alpha, \eta$ coefficients. However, due to the high complexity of the phase space structure, we haven't been able to find the corresponding eigenvalues in the most general case. Hence, in order to check the stability of the cosmological solution, we have set the parameters to the following values: $w_m=0, \alpha=-1, \eta=1, \lambda_{\phi}=1, \lambda_{\sigma}=1, z_1=1$, obtaining the eigenvalues $\approx [0.,-3.,-3.95+0.85 i,0.95\, -0.85 i,-3.95-0.85 i,0.95\, +0.85 i]$ which are associated to a saddle dynamical behavior. In this case the accelerated expansion can be associated to this cosmological solution, describing a possible early dark energy where the fields are asymptotically frozen, without kinetic energy.
\par 
Next, the $P_3^{\pm}$ critical points are located at the coordinates:
\begin{multline}
    P_3^{\pm}:=[x_1=\pm \frac{\sqrt{2} \sqrt{\eta ^2+24}}{\eta }, x_2=\frac{4 \sqrt{3}}{\eta },y_1= 0,y_2= 0,
    \\
    z_1= 0,z_2= 0],
\end{multline}
with the corresponding eigenvalues:
\begin{multline}
    \Big[ 
    0,0,3-3 w_m,\pm \frac{\sqrt{6} \alpha  \sqrt{\eta ^2+24}}{\eta }-12,3-\frac{6 \lambda _{\sigma }}{\eta },
    \\
    3\mp \frac{\sqrt{\frac{3}{2}} \sqrt{\eta ^2+24} \lambda _{\phi }}{\eta }
    \Big].
\end{multline}
These solutions are associated to a stiff--fluid case ($w_{tot}=1, s=0$) where the dark energy fluid completely dominates in terms of density parameters. As previously stated, these solutions are not of great interest for the modern cosmology, representing a not so feasible epoch driven only by the kinetic energies of the two scalar fields.
\par 
Furthermore, the $P_{4}^{\pm}$ solutions are located in the phase space structure at the coordinates:
\begin{multline}
    P_4^{\pm}:=[x_1=\pm \frac{\sqrt{2} \sqrt{\lambda _{\sigma }^2+6}}{\lambda _{\sigma }}, x_2=\frac{2 \sqrt{3}}{\lambda _{\sigma }},y_1= 0,y_2= 0,
    \\
    z_1= 0,z_2= 0],
\end{multline}
where the corresponding eigenvalues are:
\begin{multline}
    \Big[ 
    0,0,\frac{6 \eta }{\lambda _{\sigma }}-12,3-3 w_m,\pm \frac{\sqrt{6} \alpha  \sqrt{\lambda _{\sigma }^2+6}}{\lambda _{\sigma }}-12,
    \\
    3\mp\frac{\sqrt{\frac{3}{2}} \sqrt{\lambda _{\sigma }^2+6} \lambda _{\phi }}{\lambda _{\sigma }}
    \Big].
\end{multline}
As in the previous case, these stiff--fluid solutions ($w_{tot}=1, s=0$) describe an era where the expansion is not accelerated, with the dynamics driven only by the kinetic energy of the scalar fields.
\par 
The $P_5$ critical point located at:
\begin{multline}
    P_5:=[x_1=0, x_2=-\frac{\lambda _{\sigma }}{\sqrt{3}},y_1=0, y_2=\frac{\sqrt{\lambda _{\sigma }^2+6}}{\sqrt{6}},
    \\
    z_1=0 ,z_2=0 ],
\end{multline}
represents a phantom solution where the effective equation of state is equal to:
\begin{equation}
    w_{tot}=-\frac{\lambda _{\sigma }^2}{3}-1,
\end{equation}
with the density parameter of the matter component equal to zero, describing a super accelerated expansion. For this solution we have obtained the following eigenvalues:
\begin{multline}
    \Big[ -\frac{\lambda _{\sigma }^2}{2},2 \lambda _{\sigma }^2,\lambda _{\sigma } \left(2 \lambda _{\sigma }-\eta \right),
    \\
    -\frac{\lambda _{\sigma }^2}{2}-3,-\frac{\lambda _{\sigma }^2}{2}-3,-\lambda _{\sigma }^2-3 w_m-3
    \Big],
\end{multline}
corresponding to an era which is always saddle from a dynamical perspective.  
\par 
Next, the $P_6$ critical point (in the case when $w_m=0$) is located at the coordinates:
    \begin{multline*}
        P_6:=[x_1=\frac{\sqrt{3}}{\lambda _{\phi }}, x_2=\frac{2 \sqrt{3}}{\eta }, 
    \\
    y_1=\frac{\sqrt{\frac{3}{2}} \sqrt{\left(\left(\eta ^2-4\right) \lambda _{\phi }^2+\eta ^2\right) \left(5 \eta ^2 \left(5 \lambda _{\phi }^2+12\right)-366 \lambda _{\phi }^2\right)}}{\sqrt{\lambda _{\phi }^2 \left(\left(\eta ^2-4\right) \lambda _{\phi }^2+\eta ^2\right) \left(5 \eta ^2 \left(5 \lambda _{\phi }^2+12\right)-366 \lambda _{\phi }^2\right)}}, 
    \\
    y_2=0,z_1=0 ,z_2=\frac{6}{5 \eta ^2} ]
    \end{multline*}
represents a scaling solution \cite{Uzan:1999ch, Marciu:2017sji} where the effective equation of state is equal to:
\begin{equation}
    w_{tot}=w_m,
\end{equation}
with the density parameter of the matter component:
\begin{equation}
    s=\frac{132}{5 \eta ^2}-\frac{3}{\lambda _{\phi }^2}+1.
\end{equation}
For this solution we have obtained the following eigenvalues:
\begin{equation}
    \Big[\frac{3 \left(\eta -2 \lambda _{\sigma }\right)}{2 \eta },-\frac{3 \left(2 \lambda _{\phi }-\alpha \right)}{\lambda _{\phi }} ,X_3, X_4, X_5, X_6
    \Big].
\end{equation}
In this case the last eigenvalues ($X_3 - X_6$) are not written in the manuscript due to the high complexity of the formulas involved. This scaling solution is important in the phase space structure since it can explain the matter epoch. As can be noted, the kinetic energies of the two fields are set to very specific values, while the potential energy of the $\phi$ field is influenced by $\lambda_{\phi}$ and $\eta$ parameters. Moreover, the strength of the cubic coupling for the $\sigma$ field is affected by the values of the $\eta$ constant parameter in a self--interacting manner. In Fig.~\ref{fig:1} we have displayed a possible saddle region for the $P_6$ critical point by analyzing the signs of the first and second eigenvalues, considering that the first eigenvalue is positive, while the second eigenvalue is negative. As can be suspected, this analysis is not exclusive and only displays a possible region of interest associated to a saddle dynamical behavior where the two scalar fields are triggering a matter epoch, representing a scaling solution. 
\par 
The $P_7$ solution, located at the coordinates:
 \begin{multline*}
        P_7:=[x_1=0,x_2=0, y_1=0, y_2=\frac{\sqrt{2} \sqrt{\eta }}{\sqrt{2 \eta -\lambda _{\sigma }}},z_1=0 ,
        \\
        z_2=\frac{\lambda _{\sigma }}{\lambda _{\sigma }-2 \eta } ],
    \end{multline*}
corresponds to a de--Siter era ($w_{tot}=-1, s=0$) where the two scalar fields are at rest without any kinetic energy, while the dynamics is driven by the potential energy of the $\sigma$ field, and its cubic coupling term. The corresponding eigenvalues are $[0,0,-3, X_4, X_5, X_6]$, where the last three terms are not displayed due to the complicated expressions involved. As can be observed, this critical point is non--hyperbolic and can be analyzed only in the saddle behavior using the linear stability theory. For this solution, we have displayed in Fig.~\ref{fig:2} an interval associated to a saddle dynamical behavior, considering also the existence conditions which require that the $y_2$ variable is real and positive. This behavior mimics a cosmological constant dynamics and is expected to drive the evolution of the Universe in the future.
\par 
The $P_8^{\pm}$ solutions are found in the phase space structure at:
\begin{multline}
     P_8^{\pm}:=[x_1=\frac{4 \sqrt{3}}{\alpha },x_2=\pm \frac{\sqrt{2} \sqrt{24-\alpha ^2}}{\alpha },
     \\
     y_1=0, y_2=0,z_1=0 , z_2=0 ],
\end{multline}
driven a stiff--fluid scenario ($w_{tot}=1, s=0$), with the eigenvalues:
\begin{multline}
    \Big[0,0,3-3 w_m,\pm \frac{\sqrt{6} \sqrt{24-\alpha ^2} \eta }{\alpha }-12,
    3\mp \frac{\sqrt{36-\frac{3 \alpha ^2}{2}} \lambda _{\sigma }}{\alpha },
    \\
    3-\frac{6 \lambda _{\phi }}{\alpha }
    \Big].
\end{multline}

\par 
The next solution $P_9$ represents a scaling cosmological epoch ($w_{tot}=w_m$) located at the coordinates:
\begin{multline}
     P_9:=[x_1=\frac{2 \sqrt{3} \left(w_m+1\right)}{\alpha },x_2=\frac{2 \sqrt{3} \left(w_m+1\right)}{\eta },
     \\
     y_1=0, y_2=0,z_1=\frac{6 \left(w_m^2-1\right)}{\alpha ^2 \left(9 w_m+5\right)} , z_2=-\frac{6 \left(w_m^2-1\right)}{\eta ^2 \left(9 w_m+5\right)} ],
\end{multline}
with the matter density parameter equal to:
\begin{equation}
\resizebox{0.5\textwidth}{!}{$s=\frac{\alpha ^2 \left(5 \eta ^2+3 \left(3 \eta ^2+74\right) w_m-54 w_m^3+36 w_m^2+132\right)+6 \eta ^2 \left(9 w_m^3-6 w_m^2-37 w_m-22\right)}{\alpha ^2 \eta ^2 \left(9 w_m+5\right)}.$}
\end{equation}
It can be observed that for this solution the potential energy terms are absent, with the dynamics driven by the kinetic energies and the cubic components of the two scalar fields. In principle, the location in the phase space structure and the physical properties are affected by the matter component through the barotropic equation of state, and the two cubic couplings, embedded into $\eta$ and $\alpha$ parameters. For this solution we have displayed in Fig.~\ref{fig:3} a possible contour where the value of the matter density parameter is set $(s=0.3)$, obtaining various values of the associated parameters. The eigenvalues corresponding to this solution are the following (assuming a pressure--less mater component):
\begin{strip}
\begin{multline}
    \Big[-\frac{3 \alpha  \eta  \left(\alpha ^2 \left(25 \eta ^2+504\right)-504 \eta ^2\right)^2+3 \sqrt{17} \sqrt{\alpha ^2 \eta ^2 \left(\alpha ^2 \left(25 \eta ^2+504\right)-504 \eta ^2\right)^4}}{4 \alpha  \eta  \left(\alpha ^2 \left(25 \eta ^2+504\right)-504 \eta ^2\right)^2},
    \\
    \frac{3 \sqrt{17} \sqrt{\alpha ^2 \eta ^2 \left(\alpha ^2 \left(25 \eta ^2+504\right)-504 \eta ^2\right)^4}-3 \alpha  \eta  \left(\alpha ^2 \left(25 \eta ^2+504\right)-504 \eta ^2\right)^2}{4 \alpha  \eta  \left(\alpha ^2 \left(25 \eta ^2+504\right)-504 \eta ^2\right)^2},
    \\
    -\frac{3 \alpha  \eta  \left(\alpha ^2 \left(25 \eta ^2+504\right)-504 \eta ^2\right)^2+3 \sqrt{\alpha ^2 \eta ^2 \left(\alpha ^2 \left(25 \eta ^2+504\right)-504 \eta ^2\right)^3 \left(\alpha ^2 \left(425 \eta ^2+11064\right)-11064 \eta ^2\right)}}{4 \alpha  \eta  \left(\alpha ^2 \left(25 \eta ^2+504\right)-504 \eta ^2\right)^2},
    \\
    \frac{3 \sqrt{\alpha ^2 \eta ^2 \left(\alpha ^2 \left(25 \eta ^2+504\right)-504 \eta ^2\right)^3 \left(\alpha ^2 \left(425 \eta ^2+11064\right)-11064 \eta ^2\right)}-3 \alpha  \eta  \left(\alpha ^2 \left(25 \eta ^2+504\right)-504 \eta ^2\right)^2}{4 \alpha  \eta  \left(\alpha ^2 \left(25 \eta ^2+504\right)-504 \eta ^2\right)^2},
    \\
    \frac{3}{2}-\frac{3 \lambda _{\sigma }}{\eta },\frac{3}{2}-\frac{3 \lambda _{\phi }}{\alpha }
    \Big].
\end{multline}
\end{strip}
We have displayed in Fig.~\ref{fig:4} a possible region of interest for the $P_9$ solution associated to a saddle dynamical behavior which takes into account also the existence conditions.
\par 
The next critical point $P_{10}$ represents also a stiff-fluid solution ($w_{tot}=1, s=0$) found at the coordinates:
\begin{multline}
     P_{10}^{\pm}:=[x_1=\frac{2 \sqrt{3}}{\lambda_{\phi}},x_2=\pm \frac{\sqrt{2} \sqrt{6-\lambda _{\phi }^2}}{\lambda _{\phi }},
     \\
     y_1=0, y_2=0,z_1=0 , z_2=0 ],
\end{multline} with the following eigenvalues (in the case of $P_{10}^{+}$):
\begin{multline}
    \Big[0,0,\frac{6 \alpha }{\lambda _{\phi }}-12,3-3 w_m,\frac{\sqrt{6} \eta  \sqrt{6-\lambda _{\phi }^2}}{\lambda _{\phi }}-12,
    \\
    3-\frac{\lambda _{\sigma } \sqrt{9-\frac{3 \lambda _{\phi }^2}{2}}}{\lambda _{\phi }}
    \Big].
\end{multline}
In the case of a pressure--less matter component this solution cannot be stable, it is either saddle or unstable, a solution which is not very relevant from a cosmological point of view.
\par 
The $P_{11}$ solution is driven by the canonical field $\phi$,  
\begin{multline}
     P_{11}:=[x_1=\frac{\sqrt{3} \left(w_m+1\right)}{\lambda _{\phi }},x_2=0,
     \\
     y_1=\frac{\sqrt{\frac{3}{2}} \sqrt{1-w_m^2}}{\lambda _{\phi }}, y_2=0,z_1=0 , z_2=0 ],
\end{multline}
acting as a saddle matter epoch ($w_{tot}=w_m$), $s=\frac{\lambda _{\phi }^2-3 w_m-3}{\lambda _{\phi }^2}$, with the following eigenvalues: 
\tiny
\begin{multline}
\Big[\frac{3}{2} \left(w_m-1\right),-6 \left(w_m+1\right),\frac{3}{2} \left(w_m+1\right),\frac{3 \left(w_m+1\right) \left(\alpha -2 \lambda _{\phi }\right)}{\lambda _{\phi }},
    \\
    \frac{3}{4} \left(-\frac{\sqrt{\lambda _{\phi }^{10} \left(w_m-1\right) \left(\lambda _{\phi }^2 \left(9 w_m+7\right)-24 \left(w_m+1\right){}^2\right)}}{\lambda _{\phi }^6}+w_m-1\right),
    \\
    \frac{3 \left(\lambda _{\phi }^6 \left(w_m-1\right)+\sqrt{\lambda _{\phi }^{10} \left(w_m-1\right) \left(\lambda _{\phi }^2 \left(9 w_m+7\right)-24 \left(w_m+1\right){}^2\right)}\right)}{4 \lambda _{\phi }^6}
    \Big].
\end{multline}
\normalsize
At this solution the kinetic energy of the canonical field $\phi$ is influenced by the matter equation of state, and the strength of the potential energy term, encoded into the $\lambda_{\phi}$ coefficient. As can be seen, this solution is always saddle. The evolution near the $P_{11}$ critical point is represented in Fig.~\ref{fig:p11} for specific initial conditions, showing the dynamics in the $O x_1 y_1$ plane. 
\par 
Furthermore, the $P_{12}$ solution located at the coordinates:
\begin{multline}
     P_{12}:=[x_1=\frac{\lambda _{\phi }}{\sqrt{3}},x_2=0,
     \\
     y_1=\frac{\sqrt{6-\lambda _{\phi }^2}}{\sqrt{6}}, y_2=0,z_1=0 , z_2=0 ],
\end{multline}
describes a possible epoch where the acceleration of the Universe corresponds to a quintessence regime since
\begin{equation}
    w_{tot}=\frac{\lambda _{\phi }^2}{3}-1, s=0.
\end{equation}
From a dynamical point of view we have obtained the following eigenvalues:
\begin{multline}
    \Big[-2 \lambda _{\phi }^2,\frac{\lambda _{\phi }^2}{2},\lambda _{\phi } \left(\alpha -2 \lambda _{\phi }\right),
    \\
    \lambda _{\phi }^2-3 w_m-3,\frac{1}{2} \left(\lambda _{\phi }^2-6\right),\frac{1}{2} \left(\lambda _{\phi }^2-6\right)
    \Big],
\end{multline}
indicating an era which is always saddle affected in principal by the strength of the potential energy of the canonical scalar field.
\par 
Next, the $P_{13}$ critical point, located at the coordinates:
\begin{multline}
     P_{13}:=[x_1=\frac{2 \sqrt{3} \left(w_m+1\right)}{\alpha },x_2=0,
     \\
     y_1=0, y_2=0,z_1=\frac{6 \left(w_m^2-1\right)}{\alpha ^2 \left(9 w_m+5\right)} , z_2=0 ],
\end{multline}
is describing a scaling solution ($w_{tot}=w_m$) triggered by the kinetic energy of the canonical scalar field $\phi$ and the associated cubic component, with the matter density parameter equal to:
\begin{equation}
    s=\frac{5 \alpha ^2+3 \left(3 \alpha ^2-74\right) w_m+54 w_m^3-36 w_m^2-132}{\alpha ^2 \left(9 w_m+5\right)}.
\end{equation}
If we assume the case of a pressure--less matter component ($w_m=0$), then we obtain the following eigenvalues:
\tiny
\begin{multline}
    \Big[-\frac{3}{2},-6,\frac{3}{2},
    \\
    -\frac{3 \left(\alpha  \left(504-25 \alpha ^2\right)^2+\sqrt{\alpha ^2 \left(25 \alpha ^2-504\right)^3 \left(425 \alpha ^2-11064\right)}\right)}{4 \alpha  \left(504-25 \alpha ^2\right)^2},
    \\
    \frac{3 \left(\sqrt{\alpha ^2 \left(25 \alpha ^2-504\right)^3 \left(425 \alpha ^2-11064\right)}-\alpha  \left(504-25 \alpha ^2\right)^2\right)}{4 \alpha  \left(504-25 \alpha ^2\right)^2},\frac{3}{2}-\frac{3 \lambda _{\phi }}{\alpha }
    \Big],
\end{multline}
\normalsize
corresponding to a solution which is always saddle from a dynamical point of view. For this solution we have displayed in Fig.~\ref{fig:5} a specific region where the matter density parameter $s$ satisfies the existence conditions, showing the viability of such a scaling behavior.
\par 
Lastly, the $P_{14}$ solution found at the coordinates:
\begin{multline}
     P_{14}:=[x_1=0,x_2=0,
     \\
     y_1=\frac{\sqrt{2} \sqrt{\alpha }}{\sqrt{2 \alpha -\lambda _{\phi }}}, y_2=0,z_1\frac{\lambda _{\phi }}{\lambda _{\phi }-2 \alpha } , z_2=0 ],
\end{multline}
is associated to a de--Sitter epoch ($w_{tot}=-1, s=0$) where the two scalar fields are frozen. The dynamics is affected in principle by the potential energy of the canonical scalar field $\phi$, and the specific cubic component. This solution is similar to the $P_7$ critical point, representing a non--hyperbolic case where two eigenvalues are equal to zero. Since the expressions for the last eigenvalues are very complex, they are not displayed in the manuscript. As in the previous case, we show in Fig.~\ref{fig:6} a specific region of interest where such a cosmological solution is having a saddle dynamical behavior, acting closely as a cosmological constant.

\section{Summary and Conclusions}
\label{sec:3}
\par 
In the present paper we have proposed a novel dark energy model in the theoretical framework of modified gravity theories. This model extends the fundamental Einstein--Hilbert action by adding two scalar fields non--minimally coupled in an independent manner to an invariant component which contains cubic contractions of the Riemann tensor. After proposing the action for the current cosmological model, we have obtained the modified Friedmann equations by taking the variation of the action with respect to the inverse metric. The remaining field equations, the Klein--Gordon relations are developed by taking the variation with respect to the two fields which are present in the proposed action. As expected, the cosmological model satisfies the continuity equation due to the specific form of the proposed action. The dynamical features of the model are investigated by adopting the linear stability theory considering that the two fields are describing a quintom scenario, where a scalar field has a positive kinetic term, while the second field is associated to a non--canonical behavior, with a negative kinetic component, violating various physical principles from a classical point of view. From a theoretical point of view the quintom models are associated to a non--classical behavior as effective approaches in the modified gravity theories, being supported by various observations in the recent past. The analysis showed that the phase space structure is very rich in terms of stationary points, a complexity which can be associated to various stages in the evolution of the Universe. To this regard, we have obtained different classes of critical points which are associated to various cosmological eras. The first class of stationary points is represented by the stiff fluid solutions, a category which is not very interesting from a cosmological point of view. A second class of stationary points is represented by the de--Sitter solutions, a specific category which can explain the accelerated expansion where the model behaves as a cosmological constant. In this case the constant equation of state corresponds to $-1$ and can explain the late time evolution of the total equation of state. The third class of stationary points is associated to the scaling behavior where the effective equation of state corresponds to a matter epoch, a special type of solutions which can explain the existence of the matter epoch without fine--tuning. The scaling solutions found in the analysis are represented by an epoch characterized by a constant equation of state where the model behaves as a matter fluid, a dynamics driven by one or two scalar fields. In this case, the physical features are influenced by the couplings with the specific invariant which contains the third order contractions of the Riemann tensor, particular scaling solutions in the current cosmological setup. From a cosmological type of view the existence of this type of solutions in the phase space structure is very important for the viability of the corresponding models. The fourth class of solutions is associated to a quintessence or phantom behavior and can delineate the late time acceleration of the Universe by fine--tuning various parameters of the present model. These solutions can explain the quintessence--like and the phantom--like eras, as well as the possible crossing of the phantom divide line in the recent past. From the above discussion it can be noted that the present cosmological model can explain various stages in the evolution of our Universe, explaining both early and late time evolution in the phase space structure, a viable model from a theoretical point of view at the level of background dynamics.

\begin{acknowledgements}
For the development of the present manuscript we have considered analytical computations in $Mathematica$ \cite{mathematica} and $xAct$ \cite{xact}.
\end{acknowledgements}



\end{document}